\documentclass{aa}  

\usepackage{graphicx}
\usepackage{amsmath}
\usepackage{amssymb}
\usepackage{pifont}
\usepackage{float}
\usepackage{subcaption}
\usepackage{multirow}

\usepackage{txfonts}
\usepackage{hyperref}

\newcommand{\feii}{Fe\,\textsc{ii}\xspace}
\newcommand{\mgii}{Mg\,\textsc{ii}\xspace}

\begin{document}

\title{Time Evolution of \mgii in SDSS J2320+0024: Implications for a Subparsec Binary Supermassive Black Hole System}

\titlerunning{Complex  \mgii Time Evolution in SDSS J2320+0024}
\authorrunning{Fatović et al.} 

\author{Marta Fatović \inst{1,} \inst{2,} \inst{3} \href{https://orcid.org/0000-0003-1911-4326}{\includegraphics[scale=0.05]{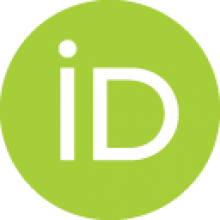}} \fnmsep\thanks{Corresponding author: marta.fatovic@unina.it} \and Dragana Ilić \inst{4,} \inst{5} \href{https://orcid.org/0000-0002-1134-4015}{\includegraphics[scale=0.05]{Orcid-ID.png}} \and Andjelka B. Kovačević \inst{4,} \inst{6} \href{https://orcid.org/0000-0001-5139-1978}{\includegraphics[scale=0.05]{Orcid-ID.png}} \and Lovro Palaversa \inst{2} \href{https://orcid.org/0000-0003-3710-0331}{\includegraphics[scale=0.05]{Orcid-ID.png}} \and Saša Simić \inst{7} \href{https://orcid.org/0000-0001-7453-2016}{\includegraphics[scale=0.05]{Orcid-ID.png}} \and Luka Č. Popović \inst{4,6,8} \href{https://orcid.org/0000-0003-2398-7664}{\includegraphics[scale=0.05]{Orcid-ID.png}} \and Karun Thanjavur \inst{9} \href{https://orcid.org/0000-0003-1187-2544}{\includegraphics[scale=0.05]{Orcid-ID.png}} \and Oleksandra Razim \inst{2} \href{https://orcid.org/0000-0002-3045-0446}{\includegraphics[scale=0.05]{Orcid-ID.png}} \and Željko Ivezić \inst{10} \href{https://orcid.org/0000-0001-5250-2633}{\includegraphics[scale=0.05]{Orcid-ID.png}}
\and Minghao Yue \inst{11,} \inst{12} \href{https://orcid.org/0000-0002-5367-8021}{\includegraphics[scale=0.05]{Orcid-ID.png}}
\and Xiaohui Fan \inst{12} \href{https://orcid.org/0000-0003-3310-0131}{\includegraphics[scale=0.05]{Orcid-ID.png}}
}

\institute{
    Dipartimento di Fisica “Ettore Pancini”, Università di Napoli Federico II, Via Cintia 80126, Naples, Italy
    \and
    Ruđer Bošković Institute, Bijenička cesta 54, 10000 Zagreb, Croatia 
         \and
         INAF – Osservatorio Astronomico di Capodimonte, Via Moiariello 16, 80131 Naples, Italy
         \and
  Department of Astronomy, Faculty of Mathematics, University of
	Belgrade, Studentski trg 16,11000 Belgrade, Serbia
 \and
 Hamburger Sternwarte, Universit{\"a}t
	Hamburg, Gojenbergsweg 112, 21029 Hamburg, Germany
 \and
 PIFI Research Fellow, Key Laboratory for Particle Astrophysics,
	Institute of High Energy Physics, Chinese Academy of Sciences,19B Yuquan Road,
	100049 Beijing, China
 \and
 Faculty of Science, University of Kragujevac, Radoja Domanovića 12, 34000 Kragujevac, Serbia
 \and
 Astronomical Observatory, Volgina 7, 11060 Belgrade, Serbia
 \and
 Department of Physics \& Astronomy, University of Victoria,
	 3800 Finnerty Road,
	  Victoria, BC V8P 5C2, Canada
   \and
   Department of Astronomy and the DiRAC Institute, University of Washington,
	 3910 15th Avenue NE,
	  Seattle, WA 98195, USA
   \and
   MIT Kavli Institute for Astrophysics and Space Research, 77 Massachusetts Ave., Cambridge, MA 02139, USA
   \and
   Steward Observatory, University of Arizona, 
933 North Cherry Avenue, Tucson, 
AZ 85721-0065, USA}

  \abstract
   {Here we present results from our spectroscopic follow-up of SDSS J2320+0024, a candidate binary supermassive black hole (SMBH) with a suspected sub-parsec separation, identified by a 278-day periodicity observed in its multi-band optical light curves. These systems may serve as a crucial link between long-period ($\sim$ tens of years) binaries, influenced by tidal forces with minimal gravitational wave damping, and ultra-short-period ($\leq$ order of days) binaries dominated by gravitational wave-driven inspiral.}
   {We investigate  the dramatic variability of the complex \mgii emission line profile aiming to test the alignments of the observed photometric light curves and the spectroscopic signatures in the context of the binary SMBH system. }
   {We extract the pure broad  \mgii line from the newly obtained Gemini and Magellan spectra and measure the emission line parameters, to reveal fundamental dynamical parameters of the SMBHs binary system. We adopt the PoSKI sub-pc binary SMBH model, which includes broad-line region around less massive component and a circumbinary broad-line region, to interpret the observed variability in the spectral profile.}
   {We find that the \mgii broad line profile has a distinctive complex shape with the asymmetry and two peaks present which is varying across recent and archival observations. The temporal variability of the \mgii line profile may be associated with the emission from the binary SMBH system consisting of components with masses \( M_1 = 2 \times 10^7 \, M_{\odot} \) and \( M_2 = 2 \times 10^8 \, M_{\odot} \), and eccentricity e = 0.1. We further discuss other plausible physical interpretations. With an total estimated mass of $\sim 10^9 M_{\odot}$ and a subannual orbital period, this system may be a rare example of high-mass compact candidate of SMBH binary,  thus important for further investigations of the evolution of the binary system. This study is a prototype of synergies of spectroscopic follow-up and future massive time-domain photometric surveys like Vera C. Rubin Observatory Legacy Survey of Space and Time.}
   {}

\maketitle
%
\section{Introduction}

Every massive galaxy is assumed to have a supermassive black hole (SMBH) at its center. When two such galaxies merge, they may form a supermassive binary black hole (SMBBH) system \citep[e.g.,][]{10.1046/j.1365-8711.2002.06056.x,2003ApJ...582..559V}. Various potential observational signatures of these binaries have been proposed, including periodic variability seen in the photometric and spectroscopic light curves and even double-peaked broad and narrow emission lines \citep[see e.g.][]{2009ApJ...705L..20X,2020MNRAS.494.4069K,2020A&A...633A..79K,2021MNRAS.505.5192P,10.1093/mnras/stab2433,Men2024}. The largest sample of possible candidates for periodically varying quasars has been identified through the analysis of large time-domain surveys. For example, \citet{Liu_2019} analysed data from the Pan-STARRS1 survey \citep[PS1;][]{2016arXiv161205560C}, initially identifying 26 candidates. However, extended observations and maximum likelihood analysis later confirmed only one statistically significant periodically varying quasar. Similarly, \citet{2020MNRAS.499.2245C} combined data from the Dark Energy Survey Supernova \citep[DES-SN;][]{2015AJ....150..172K} and the Sloan Digital Sky Survey Stripe 82 \citep[SDSS S82;][]{York_2000,Jiang_2014,Ivezic_2007}, revealing five periodically variable quasars powered by less massive black holes at high redshifts.

To enhance our understanding, it is essential to conduct follow-up observations and more detailed analyses of the most promising candidates, using a combination of techniques such as photometry across different parts of the electromagnetic spectrum, spectroscopy and direct imaging. The latter have been used to find SMBBH at kpc separations \citep[e.g., ][]{2003ApJ...582L..15K,2018ApJ...853...54S,2018ApJ...853...31O,2022A&A...658A.152V}. On the other hand, we are still in a quest to detect and confirm the presence of close binaries of SMBHs (CB-SMBHs) with sub-parsec separations, which would be on the corse to merge.

Identification of CB-SMBH's through direct methods (e.g. resolving the pair and monitoring the gas dynamics) is difficult because of the small angular separations, exceeding the resolving power of the current instruments. 
Spectroscopy, on the other hand can provide insight into the dynamics of the CB-SMBH \citep[see for example: ][ and the references within]{2017ApJ...834..129W,2023arXiv231016896D,Nguyen_2020,2012ApJ...759..118B,2016ApJS..225...29B,2012ApJS..201...23E,2013MNRAS.433.1492D,2013ApJ...777...44J,2016ApJ...822....4L,2014ApJ...789..140L,2013ApJ...775...49S,2015ApJS..221....7R,2017MNRAS.468.1683R,2017ApJ...834..129W,2019MNRAS.482.3288G, 2022MNRAS.509..212D}. Indeed, strong asymmetries and even double-peaked emission line profiles in AGNs have already been observed and studied \citep[e.g.][]{1994ApJS...90....1E,Kim_2020,2023ApJ...953L...3D}. The most thoroughly investigated cases are the ones that include changes in the line profile \citep[e.g.][]{2017MNRAS.468.1683R,refId0,Wang20,LP23}. One of the proposed explanations for this phenomenon includes two SMBH bound in a binary system. Magnetohydrodynamic (MHD) simulations provide models in which a binary system excavates surrounding material and scatters it into the circumbinary region, forming a circumbinary disk \citep[CBD,][]{2022LRR....25....3B}. The material from the inner edges of the CBD then falls onto SMBHs, forming a disk around each of them. Similar as in the case of CBD, the Broad Line Region (BLR) can be complex \citep[see][]{2021MNRAS.505.5192P}. The emission from each accretion disc continuum ionizes nearby gas, creating a broad line region corresponding to each of the SMBHs. 
Additionally, the total disc continuum emission ionizes the gas surrounding the whole system, forming a circum-binary BLR (cBLR). This setup generates broad emission lines with contributions from moving BLR1 and BLR2, and emission from the stationary cBLR. 

The recent study \citep{Fatovic_2023} reported the detection of variability in the optical light curve of five quasars located in SDSS S82 region \citep[][]{10.1093/mnras/stab1452}. The periods were calculated using the Lomb–Scargle periodogram, with the three highest periodogram peaks in the gri filters considered relevant, and only sources with gri periods consistent within 0.1\% were analyzed with the Kuiper statistic used to ensure uniform distribution of data points in phased light curves. One of the quasars (SDSS J232014.18+002459.2, hereafter SDSS J2320+0024), at a redshift z=1.05, showed a period of P=278 days which passed false alarm probability criterion. The same period was obtained by two additional independent period finding methods: i) the Quasar Harmonic Explorer (QhX) which searches for the period in quasar light curves using the cross-correlation of wavelet matrices from light curves \citep[][]{2018MNRAS.475.2051K,2019ApJ...871...32K,2020OAst...29...51K} and ii) Monte Carlo simulations with Gaussian kernel density estimation, which generated mock light curves (Tisanić et al., \textit{in prep.}), followed by period determination using the multiband Lom-Scargle periodogram implemented in \texttt{gatspy} package \citep{2015ApJ...812...18V}. Such short periods have already been found and analyzed in AGNs, for example Mrk 231 with P $\sim$ 1.1 yr, 1.2 yr \citep{Yan_2015,2020MNRAS.494.4069K}, PKS 2155-304 with P $\sim$ 0.87 yr \citep{2014RAA....14..933Z,Sandrinelli_2016}, Q J0158-4325 with P $\sim$ 172 days \citep{2022A&A...668A..77M}, and others.

\cite{Fatovic_2023} also examined the archived SDSS spectrum \citep[][]{Dawson_2016} of the SDSS J2320+0024 quasar and found a slightly asymmetrical broad \mgii emission line profile. This was the motivation to further explore this source within a spectroscopic follow-up to capture the \mgii emission line at the object's predicted maximum brightness. With this, our aim was to detect variability in the broad line profile and identify features that could point to a CB-SMBH system, such as double-peaks, peak-shifts or asymmetries.
Given the faintness of the source, 8-meter class telescopes are required to obtain high-quality spectra for a detailed analysis of the \mgii line profile.

Candidate presented in this work is different from the most extensively studied objects (e.g., PG1302-102 \citep{2015Graham}, NGC 5548 \citep{2016ApJS..225...29B}, NGC 4151 \citep{2012ApJ...759..118B}, OJ 287 \citep{1988ApJ...325..628S}) which belong to the category of systems in the early inspiral phase, with orbital periods on the order of O(10) years. Our candidate falls into the category of systems with periods on the order of O(100) days, which are approaching the late inspiral phase. These objects are very massive and exhibit significant changes in their optical spectra over very short timescales, approximately 10\% of their predicted orbital period.

In this paper we present our findings from the new observations of the  \mgii spectral line in SDSS J2320+0024, observed with Gemini and Magellan telescopes analyzed together with the archival SDSS spectrum. Furthermore, we discuss the physical meaning of the dramatic change of the double-peaked line profile and present possible model of the CB-SMBH system.

In Section 2, we present the new observations and detail the process of extracting the \mgii line. We outline the methods used to quantify the differences across the three epochs of the same line and describe the approach for mass estimation. Additionally, we explain the calculation of synthetic magnitudes and introduce the model that successfully accounts for the observed behavior of the  \mgii line. 
In Section 3, we discuss the quantified variability of the line profile and evaluate how well the synthetic magnitudes align with the predicted model. We also present  one possible physical model of CB-SMBH, emphasizing the preference for the binary model over alternative explanations. In Section 4, we place our findings within the context of other known candidates and address the challenges involved in follow-up monitoring this and similarly faint binary SMBH candidates. Finally, in Section 5, we summarize our findings and discuss potential future work.

\section{Data and Analysis}\label{sec:data_analysis}

We obtained observing time on two 8-meter class telescopes to secure two additional epochs of the \mgii spectrum. The use of 8-meter class telescopes was essential given the faintness of the target ($r_{\text{SDSS}} \sim$ 21 mag).

\subsection{Gemini observations}\label{sec:gemini_observations}
On November 14th, 2022, we obtained spectra using the Gemini Multi-Object Spectrograph-South (GMOS-S) at the Gemini Observatory (PI: Dr. Karun Thanjavur). The observations were made with longslit spectroscopy (1.0 arcseconds) using the R831 grating, during 2.2 hours of Director's Discretionary Time. We captured six science exposures, each lasting 1100 seconds, with central wavelengths alternated between 650 nm, 655 nm, and 660 nm to account for the gap between the GMOS CCDs. The Hamamatsu detector was used, and the standard star LTT3218, along with a CuAr lamp arc, provided calibration. The data were collected with a spatial binning of 4 and a spectral binning of 2. The data were reduced and calibrated using the official automated Gemini DRAGONS pipeline \citep{2019ASPC..523..321L} and cross-verified with the Image Reduction and Analysis Facility \citep[IRAF, ][]{1986SPIE..627..733T, 1993ASPC...52..173T}. We confirmed that the \mgii line is located between strong skylines, eliminating skylines as the cause of the double-peaked profile. After applying two methods for cosmic ray elimination—IRAF and the automated DRAGONS pipeline—we concluded that cosmic rays are also unlikely to be responsible for the observed line profile.

\subsection{Magellan observations}\label{sec:magellan_observations}

On December 22, 2022, observations were conducted using the Magellan 6.5-meter telescope at Cerro Pachon in Chile (PI: Dr. Xiaohui Fan), utilizing the Low Dispersion Survey Spectrograph (LDSS-3). Three consecutive integrations were performed, each lasting 1000 seconds. The setup included a VPH Blue grism and a 1-arcsecond slit, with binning set to 1x1 in both spatial and spectral directions. The slit was aligned with the average position angle (PA) of the target object. The resulting spectrum from Magellan was processed using the standard pipeline, PypeIt \citep{Prochaska2020,2020zndo...3743493P}.

\subsection{Extraction of the \mgii line} \label{sec:extraction}

In order to detect unusual features of the \mgii profile and study their behavior, it is necessary to extract the broad \mgii.
We used the Fully Automated pythoN tool for AGN Spectra analYsis \citep[Fantasy\footnote{https://fantasy-agn.readthedocs.io/en/latest/}, ][]{2023ApJS..267...19I} to subtract the underlying continuum and satellite \feii emission, following the method in \citet{2019MNRAS.484.3180P}. Fantasy provides an advanced approach to fitting multicomponent AGN spectra, allowing simultaneous fitting across a wide wavelength range and easy selection of emission lines from redefined line lists. 

\begin{figure*}[h!]
        \centering
        \includegraphics[width=1\textwidth]{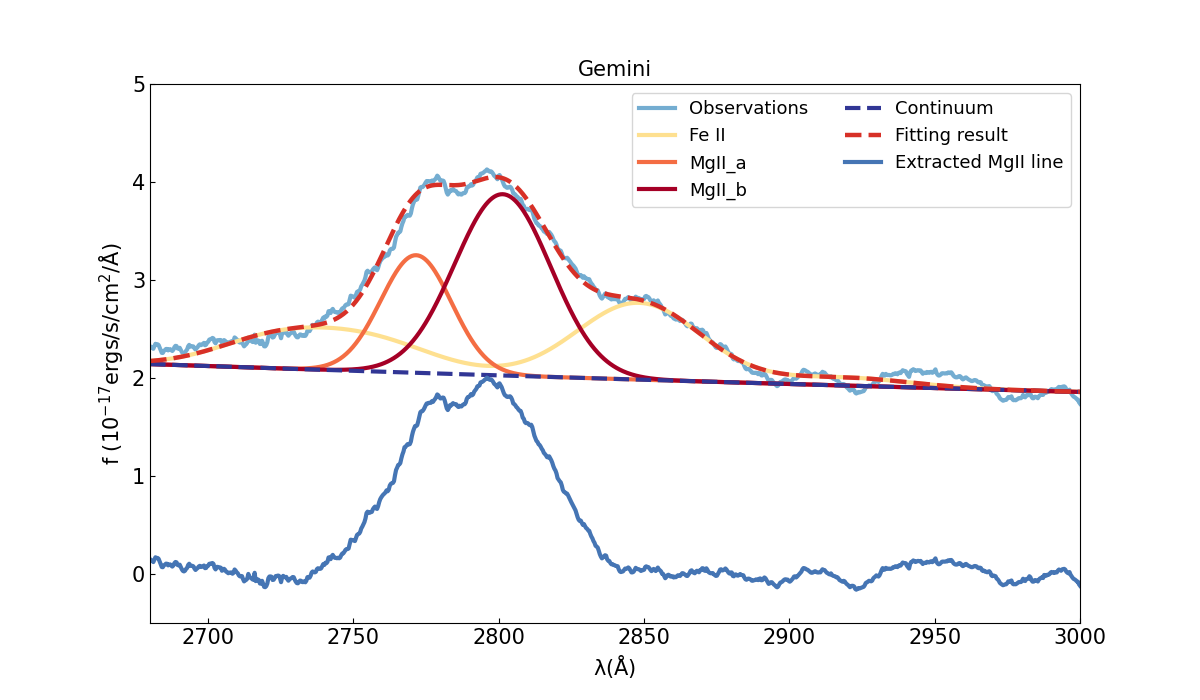}
        \label{fig:fantasy_Gemini}
    \caption{ Example of the multi-component decomposition (fitting result, dashed red line) of the complex \mgii line profile shown for observed Gemini spectrum (light blue solid line), clearly showing flat-top profile reproduced with two broad Gaussians (orange and dark red solid lines). Underlying continuum (dark blue dashed line) and Fe II multiples (yellow solid line) are also shown. Below we also show extracted pure broad profile of \mgii line (solid blue line). See text for details on the fitting procedure.}
    \label{fig:fantasy_all}
\end{figure*}

A key feature of Fantasy is its iron emission model taken from \citet{2019MNRAS.484.3180P}. To analyze the observed spectra with the distinct double-peaked shape, we fitted the broad-line region using two Gaussians while simultaneously subtracting the continuum and UV \feii emission.  
This approach minimized the $\chi^2$ value for the best fit, enabling us to effectively isolate the \mgii line for further analysis, as shown in Figure~\ref{fig:fantasy_all}. 

Using this decomposition, we successfully traced the observed flat top and complex profile features of the \mgii line with two Gaussian components, labeled \mgii\_a and \mgii\_b. These components represent the two observed \mgii peaks and are plotted in Figure~\ref{fig:fantasy_all}, along with the underlying continuum and the \feii multiplets. It is worth noting that we were able to extract the SDSS \mgii line using two Gaussians, despite it being the only observation without a clearly defined double-peaked profile.

\subsection{Characterizing Line Profile Variations} \label{sec:changes_meas}

After extracting the \mgii line, we proceeded with an analysis using a set of well-established, easily measurable line parameters to capture and describe the complex variations in the line profile \citep[e.g., see the approach outlined in ][]{2010ApJS..187..416L}.
To accurately characterize the  \mgii line only, we fitted the extracted broad  \mgii profile using two Gaussian components. From these new fits, we measured several key parameters \citep{2019MNRAS.484.3180P}. 
These parameters include: i) velocity shifts of the two peaks from their rest positions, ii) intensity ratio of the red and blue peaks, iii) full width at half maximum (FWHM), iv) full width at quarter maximum (FWQM), v) the distance between the red and blue peaks, and vi) profile asymmetry (skewness) which is measured by comparing the shifts between the center of the profile at both half-maximum (Shift$_{\rm HM}$) and quarter-maximum (Shift$_{\rm HM}$) with the zero velocity reference point (0 km s$^{-1}$). The findings are listed in Table~\ref{tab:meas}.

\subsection{Mass and separation estimation} \label{sec:mass_est}
We calculate a first estimate of the total mass of the system from the width of the broad  \mgii
line and continuum flux at 3000$\AA$ using the standard scaling relation given in \citet{Wang_2009, 2013AA...555A..89M, Popovic+2020+1+14}:
\begin{equation} \label{eq:mass_estimation}
\log(M_{\mathrm{BH}}) = a + b \cdot \log(L_{\lambda3000}) + c \cdot \log(\mathrm{FWHM}_{\mathrm{ \mgii}}),
\end{equation}
{with constant values: $a = 1.15 \pm 0.27$, $b = 0.46 \pm 0.08$, and $c = 1.48 \pm 0.49$, where the mass is given in $10^6 M_\odot$, where \( M_{\odot} \) denotes the solar mass, the $\mathrm{FWHM}_{\mathrm{ \mgii}}$ in $10^3$ km s$^{-1}$, and the $L_{\lambda3000}$ in $10^{44}$ erg s$^{-1}$.
The formal errors are determined from independent measurements, using different estimates of the underlying continuum, which is primarily influenced by the spectral resolution. 

We also calculate the separation of the black holes in a possible SMBBH system via Kepler's law as in \citet{Liu_2019}:
\begin{equation} \label{eq:Kepler}
    \frac{a^3}{t_{\text{orb}}^2} = \frac{GM}{4\pi^2},
\end{equation}
where
\begin{equation} \label{eq:t_orb}
    t_{\text{orb}} = \frac{P_{\text{obs}}}{1 + z}.
\end{equation}

In all our analyses, we used the cosmological parameters provided by \cite{2020AA...641A...6P}.

\subsection{Synthetic magnitude calculation} \label{sec:synth_mags}

In order to compare the model light curve with the observed data, we calculated synthetic SDSS \textit{r}-band magnitudes from Gemini and Magellan spectra. For obtaining synthetic magnitudes in other SDSS bands, the wavelength coverage of these spectra is not broad enough. For deriving synthetic fluxes we used the \texttt{rubin\_sim} package \citep{peter_yoachim_2023_8388546}, which is being developed and maintained by the Legacy Survey of Space and Time \citep[LSST; ][]{2019ApJ...873..111I} community and which already contains SDSS throughputs and zero points. The calculation of the synthetic flux is done by convolving the flux of the spectra with the survey's throughput functions (in our case, SDSS \textit{r}-band) and then by integrating it:
\begin{equation}
	F_b=\int\limits_0^\infty \phi_b(\lambda)F_{\nu}(\lambda)\,\mathrm{d}\lambda
\end{equation}
where $F_b$ is the flux in a band $b$, $\phi_b(\lambda)$ is the bandpass throughput and $F_{\nu}(\lambda)$  is the observed flux density of a source.
The AB magnitudes are then calculated using $m_b = -2.5 \cdot \log_{10}F_b - C_b$, where $\mathrm{C_b}$ is the zero-point.

\subsection{PoSKI model of binary SMBH system} \label{sec:poski_method}

Finally, given the detected variability in the emission line profile that could be indicative of the complex dynamics within the system, we attempt to model complex \mgii broad line in the context of binary SMBH system using the Popović, Simić, Kovačević, Ilić model \citep[PoSKI, ][]{2021MNRAS.505.5192P}. In the PoSKI model, two SMBHs at a sub-pc distance each have accretion discs that ionize nearby gas, forming two moving broad-line regions (BLR1, BLR2). The combined disc emission also ionizes gas around the system, creating a stationary circum-binary broad-line region (cBLR). This produces broad emission lines from three sources: BLR1, BLR2, and the cBLR. For the details of the model, see \cite{2021MNRAS.505.5192P} and the references therein.

\section{Results}\label{sec:results}
\begin{figure*}[h!]
\centering
\includegraphics[width=0.85\textwidth]{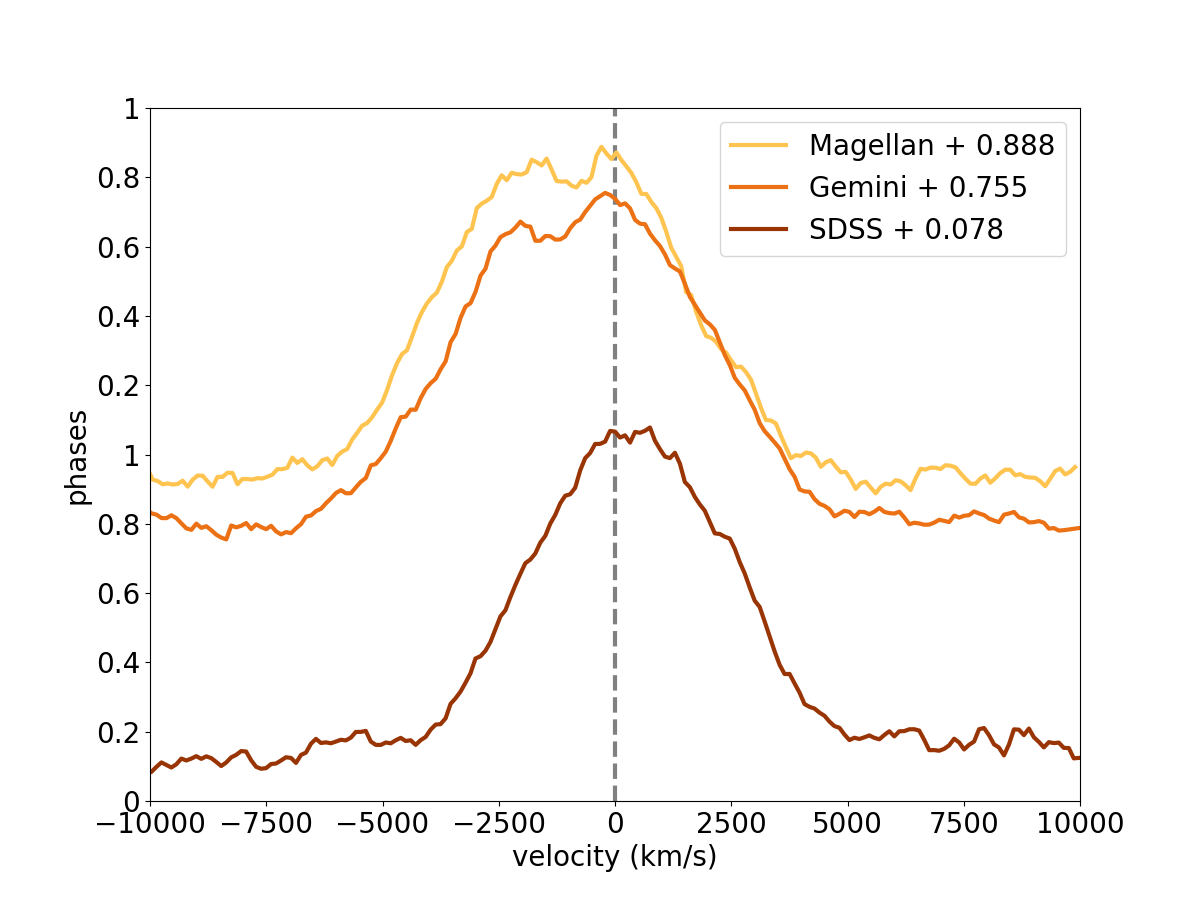}
\caption{The normalized line profiles of the extracted  \mgii line, as observed with SDSS (brown solid line), Gemini (orange solid line), and Magellan (yellow solid line) telescopes, sorted by the phase derived from presumed periodic variability of the optical light curves. The grey vertical dashed line indicates a velocity of 0 km s$^{-1}$.}\label{fig:spec}
\end{figure*}
\begin{figure*}[h]
\centering
\includegraphics[width=0.8\textwidth]{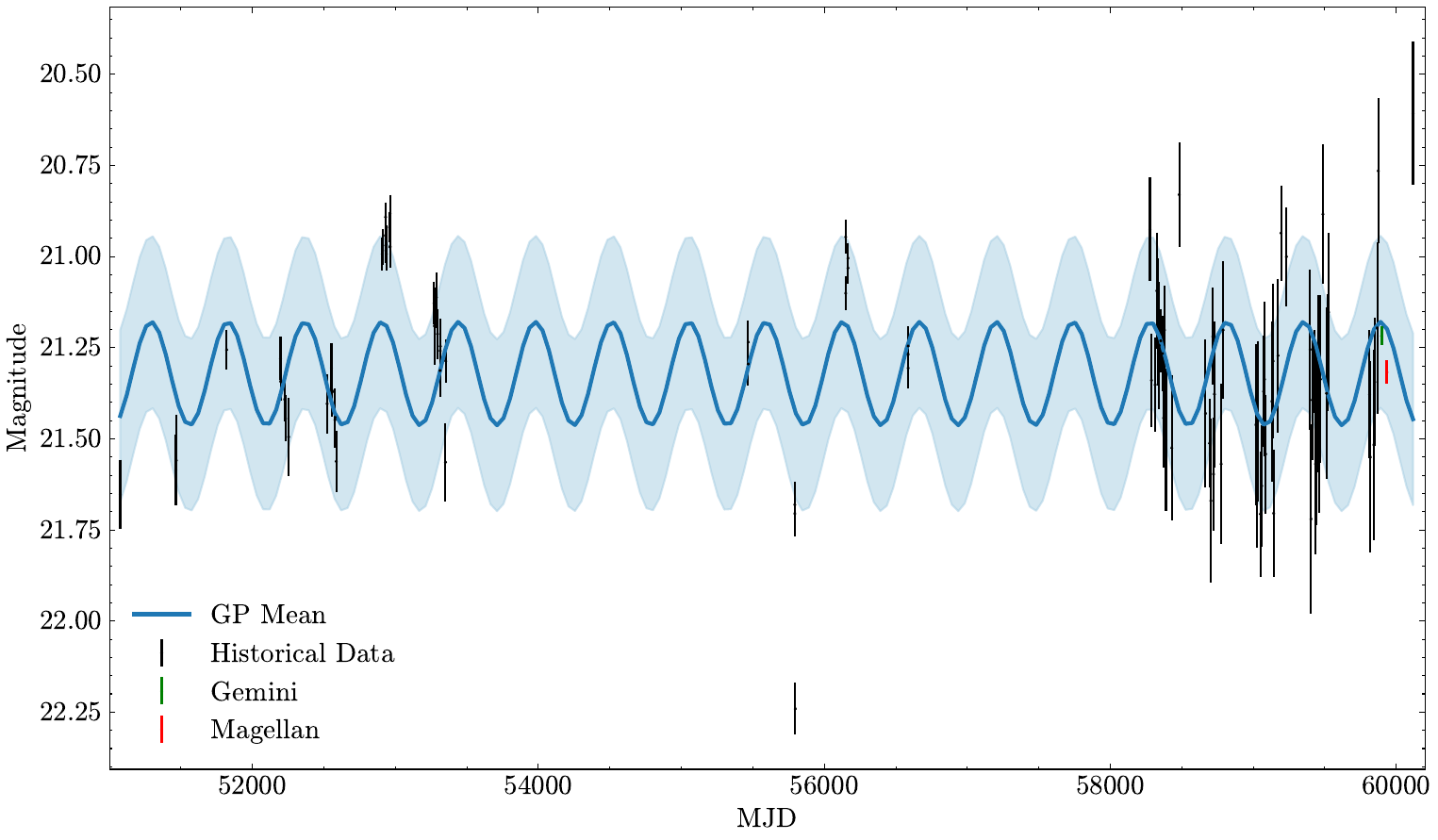}
\caption{Light curve of the SDSS J2320+0024 source in the \textit{r} band (black error bars) of historical SDSS, ZTF  and PS1 data. GP periodic curve (blue  line) that best matches the data, shows a period of $\sim$270 days and an amplitude of 0.3 mag. The blue lent  represents the 1$\sigma$ uncertainty of the GP curve. Green and red error bars represent the newly observed Gemini and Magellan \textit{r} magnitudes, respectively,  extracted using our procedure (see text).  GP curve  highlights the placement of Magellan and Gemini observations within $1\sigma$ of the expected waveform of binary orbital period. }\label{fig:LC}
\end{figure*}

\subsection{Line profile variability} \label{sec:measurements}
Dramatic evolution of \mgii broad emission line profile displayed in velocity scale and shifted according to the observed phase based on photometric light curve in \citet{Fatovic_2023}, is shown in Figure \ref{fig:spec}.
The peak separation changed for $\sim$400 km s$^{-1}$, and the peak intensity ratio changed from 0.893 to 0.973 within the month that had passed between Gemini and Magellan observations. SDSS profile displays a single peak corresponding to the position of the red peak in the other two profiles.

There is an approximate 1000 km s\(^{-1}\) difference in FWHM, FWQM, and asymmetry/skewness (see Table~\ref{tab:meas}) between the SDSS line profile and the profiles in the new spectra. Notably, the asymmetry/skewness measured at half and quarter maximum (Shift\(_{HM}\) and Shift\(_{QM}\)) shows a significant change of about 600 km s\(^{-1}\) and 300 km s\(^{-1}\), respectively, between the Magellan and Gemini observations taken within a one-month interval. The red peak remains steady while the blue peak shifts, changing the overall shape of the \mgii line profile. This observation inspired our efforts to model and explain these variations with the PoSKI model of a CB-SMBH system.

The results of the measurements of line parameters for all three epochs are given in Table~\ref{tab:meas}. 
The system's mass estimation for all three spectra is $M_{\scalebox{0.5}{\text{\ding{108}}\text{\ding{108}}}} \sim 10^{9} \, M_{\odot}$. This makes SDSS J2320+0024 one of the most massive sub-annual binary quasar candidates known.

\begin{table*}[h!]
\begin{centering}
\resizebox{0.6\textwidth}{!}{
\begin{tabular}{cccc}
\hline
 & SDSS & Gemini & Magellan \\
 \hline
log($M_{\scalebox{0.5}{\text{\ding{108}}\text{\ding{108}}}} / M_{\odot}$\,) & 9.0 $\pm$ 0.5 & 8.9 $\pm$ 0.5 & 8.8 $\pm$ 0.5 \\
Shift$_{B}$ (km s$^{-1}$) & NA & -1801 $\pm$ 107 & -1453 $\pm$ 71 \\
Shift$_{R}$ (km s$^{-1}$) & 356 $\pm$ 71 & -140 $\pm$ 107 & -125 $\pm$ 71 \\
$I_{B}/I_{R}$ & NA & 0.893 $\pm$ 0.004 & 0.973 $\pm$ 0.004 \\
FWHM (km s$^{-1}$) & 5254 $\pm$ 71 & 6219 $\pm$ 107 & 6112 $\pm$ 71 \\
FWQM (km s$^{-1}$) & 7291 $\pm$ 179 & 8470 $\pm$ 250 & 8256 $\pm$ 500 \\
Peak separation (km s$^{-1}$) & NA & 1661 $\pm$ 107 & 1328 $\pm$ 71 \\
Shift$_{\rm HM}$ (km s$^{-1}$) & 377 $\pm$ 36 & -640 $\pm$ 36 & -1203 $\pm$ 38 \\
Shift$_{\rm QM}$ (km s$^{-1}$) & 324 $\pm$ 89 & -694 $\pm$ 36 & -988 $\pm$ 250 \\
a (pc) & 0.004 $\pm$ 0.001 & 0.003 $\pm$ 0.001 & 0.003 $\pm$ 0.001 \\
\hline
\end{tabular}}
\caption{Measured quantities from the broad  \mgii emission line observed with SDSS, Gemini and Magellan. Rows: log($M_{\scalebox{0.5}{\text{\ding{108}}\text{\ding{108}}}} / M_{\odot}$\,): logarithm of the estimated system's mass given in solar masses; Shift$_{B}$ and Shift$_{R}$: peak shift of the blue and red peak in relation to the 0 km/s; $I_{R}/I_{B}$: ratio of the intensities of red and blue peaks; FWHM and FWQM: width of the whole profile at 50\% and 25\%; Peak separation: velocity separation of the blue and red peaks; Shift$_{HM}$ and Shift$_{QM}$: velocity shift of the profile centroid at 50\% and 25\%; a: the separation.}
\label{tab:meas}
\end{centering}
\end{table*}

\subsection{New synthetic magnitudes in the photometric light curves} \label{sec:synth_lc}

We calculated the synthetic magnitudes from the spectroscopic observations because we wanted to compare how the variations in the line profile correlate with the photometric variations. Figure \ref{fig:LC} shows a light curve, reproduced by a mix of observed (SDSS, PS1, Zwicky Transient Facility \citep[ZTF; ][]{2019PASP..131a8002B,2019PASP..131g8001G}) and synthetic photometry obtained by the Gemini and Magellan, overlaid on the sinusoidal model of periodic variations of SDSS J2320+0024. Since the ZTF data showed a linear trend in the light curve, we preformed linear detrending. 

For the subsequent light-curve modeling, we relied solely on observational data from SDSS, PS1, and ZTF. This approach provided an additional confirmation of the results reported in \cite{Fatovic_2023}, this time using a different method and an extended dataset spanning over 20 years, incorporating observations from ZTF, PS1, and SDSS. To model a quasar light curve, we used the \texttt{GPyTorch} Python package \citep{gardner2018gpytorch}, applying a Gaussian Process (GP) with a kernel defined as:
$$
K(x, x’) = \sigma^2 \cos\left(\frac{2\pi |x - x’|}{P}\right),
$$
where $\sigma^2$ is the signal variance (a scaling factor), $P$  is the period (a learnable parameter), and $x, x';$ are input points. We initialized the model’s mean function with the observed mean magnitude of the light curve. The \texttt{CosineKernel} was chosen for its ability to model periodic behavior and was initialized with an approximate period of 265 days, which was set as a learnable parameter to enable fine-tuning based on the data. During training, we used an Exact Marginal Log Likelihood (MLL) objective to optimize both the kernel parameters and the GP’s likelihood function with an Adam optimizer. After training, the model was evaluated over a test range of 200 points randomly selected from the observed baseline, providing a predicted mean and confidence intervals for each time point. The results revealed a fitted periodic pattern with learned confidence intervals, and the model successfully estimated the period at approximately 270 days.

The synthetic magnitudes from Gemini and Magellan fall within the 1$\sigma$ confidence interval of the modeled GP light curve. The SDSS synthetic magnitude deviates by approximately 3$\sigma$, and it has been intentionally excluded from Figure~\ref{fig:LC}. We opted not to use SDSS fluxes because of the uncertainties related to the absolute flux calibration \footnote{https://www.sdss4.org/dr16/algorithms/spectrophotometry/}.

\subsection{Implementation of the PoSKI model} \label{sec:PoSKI_implementation}

\begin{figure}[h!]
\centering
\includegraphics[width=0.4\textwidth]{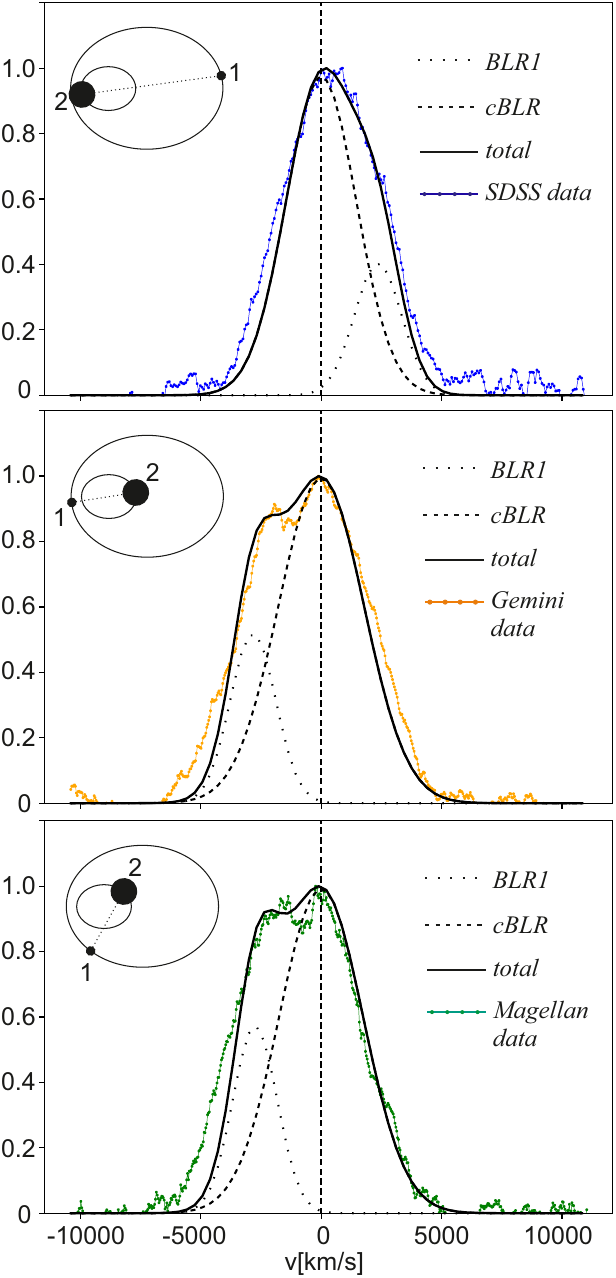}
\caption{PoSKI model of  \mgii broad line (black solid line) for all three epochs (top panel: SDSS, middle panel: Gemini, bottom panel: Magellan data). The components are coming from the smaller BLR (BLR1, dotted line) and cBLR (dashed line). Upper insets in each panel shows the binary configuration for the corresponding epoch. The wavelength is converted to the velocity scale given on the x-axis, whereas the y-axis shows the normalized intensity.}\label{fig:config}
\end{figure}

Due to the dramatic changes in the position of the blue peak in the complex \mgii line profile (see Figure \ref{fig:spec}) which appear to be correlated to the light curve (see Figure \ref{fig:LC}) indicating possibility of a binary system, we attempted to use PoSKI model to interpret the observed behavior. To find a model of SMBBHs that can describe the observed variability and complex  \mgii line shape, we explored several different configurations of SMBBHs, changing the mass ratio and dynamical parameters but fixing the periodicity. This observations motivated us to set up the PoSKI model in a configuration with two SMBHs with a mass ratio q$\sim$0.1, where only the less massive component has a broad-line region (BLR1) contained within its Roche lobe \citep[see][]{2021MNRAS.505.5192P}, moving with the component. Both components have accretion discs that illuminate the gas around both SMBHs, which is circum-binary BLR. We assumed that the more massive component did not have enough ionized gas in the Roche lobe to have its own BLR. It may be the case since smaller component cleans the material around the more massive component. Using this configuration of SMBBH, we are able to qualitatively fit the observed \mgii line profile and reproduce the observed periodicity in the continuum.

Optical photometry allowed us to determine the orbital period of the object, which allowed us to place broad constraints on the component masses and the distance. Presumed orbital velocities of such a system would be too high and would produce significant line shifts, not supported by spectroscopic observations. Therefore, we  used in our modelling the inclination angle of very low value \citep[$i = 10^{\circ}$, which has been also implemented in previous theoretical works, see e.g.,][]{Wang_2018} and compute the emission from the SMBBH. Additionally, observations of Gemini and Magellan are very close in time, and the light curve variation between those time instances allows us to put additional constraints on the eccentricity and orbital plane orientation toward the observer. Since  \mgii line deviation is asymmetric, we propose the low mass ratio binary system, which has the mass-ratio parameter $q=0.1$.

The parameters used in the simulation define the physical properties and orbital configuration of the system. The suggested masses of the two SMBHs are \( M_1 = 2 \times 10^7 \, M_{\odot} \) and \( M_2 = 2 \times 10^8 \, M_{\odot} \), indicating the less massive and more massive components. The mean separation between the two SMBHs is \( a = 0.0025 \, \text{pc} \). Furthermore, the orbital eccentricity, \( \text{e} = 0.1 \), reflects a mildly elliptical orbit, while the inclination of the orbital plane relative to the line of sight is \( \text{i} = 10^\circ \). The BLR in such a compact case is truncated due to the mutual interaction of the components. The computational phases, expressed as fractions of the orbital period (\( P_{\text{orb}} \)), are \( t_{\text{SDSS}} = 0.02 \times P_{\text{orb}} \), \( t_{\text{Gem}} = 0.53 \times P_{\text{orb}} \), and \( t_{\text{Mag}} = 0.7 \times P_{\text{orb}} \).

The three suggested binary SMBH configurations and resulting \mgii broad line profiles from PoSKI models are plotted in Figure~\ref{fig:config}. The contribution from BLR1 (associated with the less massive SMBH) is shown as a dotted line, while the cBLR contribution is represented by a dashed line.

\section{Discussion} \label{sec:discussion}

The results of our study reveal a candidate CB-SMBH system in SDSS J2320+0024 characterized by a total mass of \( \sim 10^9 \, M_{\odot} \), a mean separation of \( \sim 0.0025 \, \mathrm{pc} \), and an orbital period of 278 days. These properties place it among the most compact and dynamically extreme candidate CB-SMBH systems known. Predictions by \citet{2021MNRAS.506.2408X} suggest the upcoming LSST data will offer optimal opportunities for detecting these sources. Accurate light curves in \textit{ugrizy} filters and a limiting LSST magnitude of $\sim$ 24 mag from single image will be of exceptional value. When comparing our source (\( z = 1.05 \), \( r_{\text{SDSS}} \sim 21 \) mag, \( P = 278 \, \text{ days} \)) to Figure 7 in \citet{2021MNRAS.506.2408X}, we note that they consider periods up to 200 days. This places our source near the upper bound of the predicted binaries, corresponding to a density of approximately \( \log(N[\text{mag}^{-1} \, z^{-1} \, \text{day}^{-1}]) \sim 1 \). 

\citet{2023MNRAS.526.1588} reported CB-SMBH candidates with orbital periods as short as 340 days, while \citet{2022A&A...668A..77M} identify systems with periodicities around 173 days, both broadly consistent with the timescales of the source presented here. However, spectroscopic monitoring and analysis of variability in spectral lines offers important comparisons. Specifically, \citet{2022A&A...668A..77M} presented the \mgii line profile of their candidate, which is symmetric with a prominent peak. However, the spectrum of this candidate used in \citet{Faure} could be interpreted as asymmetric, though no detailed analysis was performed. In contrast, the \mgii broad line profiles of the object in this study exhibit distinct features in each observed epoch. The sources discussed by \citet{2023MNRAS.526.1588} exhibit double-peaked broad emission lines, such as H\(\alpha\), which are attributed to Doppler shifts caused by orbital motion. Similarly, \cite{2009Natur.458...53B} analyzed the source J153636.22+044127.0, which significantly contributed to studies of SMBBH systems by presenting two distinct sets of Balmer broad lines. However, \cite{2019ApJ...877...33Z} later suggested that double sets of emission lines may often be false-positive indicators of black hole binaries. Instead, such features could result from alternative physical mechanisms, such as AGN-driven or shock-heated outflowing gases, which produce blueshifted broad emission-line systems.  

In contrast, the system in our study stands out due to its variability, first identified through photometric observations and later obtained the same result within \(1\sigma\) using spectroscopic data from the broad \mgii line. The significant variability in the \mgii line profile suggests that the photometric variability cannot be attributed solely to red noise, pointing to dynamic processes like binary supermassive black holes. 

This configuration generates a signal that should be detectable in all broad emission lines, with the shape of the complex line profile varying depending on the system's specific dynamical configuration. Therefore, to more reliably support the hypothesis for a CB-SMBH in SDSS J2320+0024, and to explore alternative explanations such as a single black hole model with complex BLR kinematics, the analysis should include additional emission lines, such as H\(\beta\), H\(\alpha\) and others. Furthermore, testing the model across multiple epochs and conducting dedicated photometric monitoring would be essential for detecting variability consistent with binary motion.

Our 2385 $\mathrm{s}$  Swift UVOT integration carried out on 11 October 2023 through UVW1 \citep{2004ApJ...611.1005G} resulted in no UV detection above S/N > 3 threshold. Particularly this source is too faint even for SWIFT XRT. We stress that the faintness of the source (SDSS J2320+0024, $r_{\rm SDSS}\sim21$ mag) and its limited visibility during the year — observable from ground-based facilities only from August to late December - further complicate follow-up with the goal of building up an entire phased light curve. These constraints make tracking this and similar faint CB-SMBH systems particularly challenging, especially in ground-based surveys where atmospheric effects, noise, and blending issues hinder variability analyses. However, advancements such as LSST’s high cadence and deep imaging capabilities \citep{2024ApJ...965...34D} offer a promising pathway to detect such faint, short-period systems. Additionally, the role of space-based telescopes, like the James Webb Space Telescope \citep[JWST; ][]{2006SSRv..123..485G}, with their superior sensitivity and broad coverage, remains crucial for studying these systems.

\subsection{Complex \mgii within the context of binary SMBH} \label{sec:model_disc}

To reproduce the complex behavior observed in the data within the binary SMBH framework, we applied PoSKI model that can provide dynamical parameters for close-binary systems.
Our analysis shows that with the suggested binary configuration (see Sec. \ref{sec:PoSKI_implementation}), the PoSKI model is able to consistently reproduce \mgii broad line profiles in all three observational epochs. However, the slight discrepancies are still seen, especially in blue wing ($\sim-5000$ $\mathrm{km/s}$, Figure~\ref{fig:config}) in case of all epochs. 

These discrepancies might be accounted on the observed data quality and later analysis, which is influencing the extraction of the pure broad line profile. E.g., the accurate reconstruction and subtraction of the \feii emission, that may be having complex variability and kinematical origin, may have a strong influence on the resulting broad line profile. Also, these may be an evidence of the presence of non radial motions in the \mgii emitting region \citep[see e.g.,][]{2019MNRAS.484.3180P, 2019ApJ...883L..44G,2020MNRAS.496..309H, 2024ApJS..270...26G}, such as inflows or outflows. 

\subsection{Possible alternative interpretations} \label{sec:why_binary_model}

Alternative explanations for the periodic variability observed in the optical light curve \citep[see][]{Fatovic_2023} include jet precession and warped accretion discs. Sources with powerful jets are typically expected to appear in radio databases and show strong variability in radio periodicity \citep[see discussion in e.g. ][]{2015Graham}. However, there is little evidence for radio emission from this source, as it was not detected in several radio surveys, including the Very Large Array (VLA) FIRST Survey at 1.4 GHz \citep{becker}, the AT20G Survey at 20 GHz \citep{Massardi}, or the first two epochs of the VLA Sky  Survey. Additionally, it is absent from the 1.4 GHz catalog by \citet{Hodge_2011}, which offers slightly higher sensitivity (0.09 mJy compared to 0.13 mJy). While the lack of radio detections does not entirely rule out jet precession as the cause of the observed variability, it makes this explanation less likely. \cite{2013ApJ...763L..36L} showed that the \mgii line can be formed in the jet-like structures and that its variability is highly correlated with the gamma and radio emission. However, in their object the broad \mgii line profile remains to be symmetric and there is only a change in the luminosity. This is not the case in our object in which there is clear change in the line profile, indicating either significant perturbation in the BLR structure and kinematics or the presence of additional BLR component.
 
Some studies suggest that warped disks can change emission line profiles depending on their geometry \citep[e.g., ][]{1997ApJ...489...87S,Wu,2012ApJ...744..186W}. This implies that, from a spectroscopic perspective, the candidate in this work could be associated with the precession of a warped disk. From a photometric perspective, warped disks are predicted to produce quasi-periodic oscillations (QPOs) in X-ray light curves \citep[e.g., ][]{Abarr_2021}. While QPOs have been detected in optical light curves \citep[e.g.,][]{Smith_2018, tess}, our optical data point to periodic variability rather than QPOs. Furthermore, light curves associated with known warped disks typically display lower amplitude variations and distinct behaviors compared to our observations \citep[see, e.g.,][and references therein]{2015Graham}. This suggests that the observed photometric periodicity is less probable to originate from a warped disk.

Furthermore, double-peaked emission lines have been explored in previous studies, such as \cite{2023ApJ...953L...3D}, where simultaneous double peaks were observed in two broad emission lines (O I and Pa$\alpha$) and attributed to a disk-like BLR. Notably, however, their analysis did not report variability in the line profiles, a significant contrast to our findings, thereby making this explanation unlikely for our observations.

Temporal changes in the \mgii line asymmetry have also been observed in some AGNs, such as J111348.6+494522, as reported in \cite{2020MNRAS.496..309H}. They explain these changes as being caused by variations in the relative intensities of line components, which may represent the broad and narrow features or reflect a more dynamic BLR structure. This explanation aligns with some alternative scenarios, such as a single black hole model with complex BLR kinematics \citep[e.g. ][]{Grier_2017,2019A&A...621A..46E,Rodriguez-Ardila_2024} or flaring \citep{Chavushyan_2020}, remain plausible. For example radial motions within the BLR such as outflows may be responsible for the observed line asymmetry \citep[see e.g. ][]{2017NatAs...1..775W}, however it is challenging to explain this in the frame of periodic variability on observed timescales. To further explore this possibility, future analyses should incorporate additional spectral observations with top-level instruments of \mgii as well as other emission lines, such as H\(\beta\), H\(\alpha\), and others.

Binary model suggested in this work may offer a more plausible explanation, supported by photometric, and in this work, even with spectroscopic evidence. From spectral analysis (Sections \ref{sec:changes_meas} and \ref{sec:mass_est}; results listed in Table~\ref{tab:meas}), we estimate a total system mass on the order of $10^9$$M_\odot$, separations between the two components on the order of milliparsecs, and extremely changing broad line profile. These results align with predictions from PoSKI model (see Section~\ref{sec:PoSKI_implementation} and Figure) and the observed light curve periodicity, which corresponds to luminosity changes driven by the orbital motion of the components. Also, based on the PoSKI model parameters, measured 1000 ${\rm km/s}$ broadening reflects gas near the inner edge of the circumbinary disk ($\sim2\times0.003$pc). Additionally, observed 600 ${\rm km/s}$ at FWHM (closer to SMBH) in line core and 300 ${\rm km/s}$ (FWQM, further) asymmetries may arise from the dynamics of gas in the circumbinary disk or from the BLR around the less massive component. Contrary, if  single SMBH assumed then plausible asymmetry changes would not dominate the line core (as it is now 600 ${\rm km/s}$), but mostly line wings.  

\section{Conclusion}\label{sec:conclusion}

In this study, we report the results of spectroscopic follow-up conducted with 8-meter class telescopes of SDSS J2320+0024, a sub-parsec binary SMBH candidate, identified through its distinct periodicity in photometric multi-band optical light curves, suggestive of a binary orbital period of 278 days \citep{Fatovic_2023}. Such systems are critical for understanding the transitional dynamics as they approach the gravitational waves (GW) detection threshold, particularly within the operational parameters of the Laser Interferometer Space Antenna \citep[LISA; ][]{2017arXiv170200786A}. It is expected that GW damping in these binaries, while significant, might not yet predominate the accretion processes. Moreover, these binaries constitute a relatively sparse population within current catalogs.

We analyze the variability of the broad \mgii emission line across three epochs—archival SDSS and new Gemini and Magellan spectra. We also check how the synthetic magnitudes obtained from the Gemini and Magellan spectra follow the long-term photometric light curve. We compare the observed spectra from each epoch with the PoSKI model of a CB-SMBH system with unequal masses. 

We summarize our findings as follows:  

\begin{enumerate}[(i)]
    \item We report dramatic variability in the complex broad \(\text{Mg II}\) emission line profile, observed during the spectroscopic follow-up.
    \item Our analysis estimates the total mass of the system at \(\sim 10^9 M_{\odot}\), with significant shifts of several hundred \(\mathrm{km \, s^{-1}}\) between the two peaks. This substantial mass places the object among the most massive subannual binary quasar candidates known, suggesting a history of extensive galaxy mergers. 
    \item The spectra were successfully interpreted within the framework of a binary SMBH system using the PoSKI model. 
    \item The orbital period inferred from the PoSKI model aligns with the periodicity identified in the historical photometric light curve. 
    
\end{enumerate}

    If confirmed through spectroscopy of additional emission lines, these results could provide valuable insights into the capabilities of upcoming large-scale time-domain optical surveys and lay the foundation for future multimessenger studies \citep{2022MNRAS.510.5929C}. Spectroscopic surveys accompanying such observations are expected to identify and analyze massive binary quasars, significantly enhancing our understanding of galaxy merger rates. This study outlines a pathway for future investigations, emphasizing the synergy between the Rubin Observatory LSST survey and spectroscopic follow-ups with advanced instruments, such as those planned for the Wide-field Spectroscopic Telescope \citep[WST; ][]{2024arXiv240305398M}, Multi-Object Optical and Near-infrared Spectrograph \citep[MOONS; ][]{2020Msngr.180...10C},  4-metre Multi-Object Spectroscopic Telescope \citep[4MOST; ][]{2019Msngr.175....3D}, WHT Enhanced Area Velocity Explorer \citep[WEAVE; ][]{2024MNRAS.530.2688J}, Maunakea Spectroscopic Explorer \citep[MSE; ][]{2019clrp.2020...30H}, and others.

\begin{acknowledgements}
    We thank Gemini North Observatory for hosting M.F. and L.P. and for providing valuable guidance on data reduction. Special thanks to German Gimeno, Kathleen Labrie, and Christopher Simpson for their assistance in the Gemini data reduction process. We thank Amy Kimball, Nicole Nesvadba, Romain Petrov and James Leftley for helping us with the observing proposals. This work is financed within the Tenure Track Pilot Programme of the Croatian Science Foundation and the Ecole Polytechnique Fédérale de Lausanne and the Project TTP-2018-07-1171 Mining the variable sky, with the funds of the Croatian-Swiss Research Programme. D.I., A.B.K., L.\v C.P.  and S.S. acknowledge funding provided by the University of Belgrade - Faculty of Mathematics (the contract 451-03-47/2023-01/200104), Astronomical Observatory Belgrade (the contract 51-03-47/2023-01/200002) and the University of Kragujevac-Faculty of Sciences (the contract 451-03-68/2022-14/200122) through the grants by the Ministry of Science, Technological Development and Innovation of the Republic of Serbia. D.I. acknowledges the support of the Alexander von Humboldt Foundation. A.B.K. and L.\v C.P. thank the support by Chinese Academy of Sciences President’s International Fellowship Initiative (PIFI) for visiting scientists. MF acknowledges the ﬁnancial contribution from PRIN-MIUR 2022 and from the Timedomes grant within the “INAF 2023 Finanziamento della Ricerca Fondamentale”.
\end{acknowledgements}

\bibliographystyle{aa} 
\bibliography{aanda}

\end{document}